# Satellite Optical Brightness

Forrest Fankhauser[1], J. Anthony Tyson[1], and Jacob Askari[2]
[1] Department of Physics and Astronomy, University of California, One Shields Avenue, Davis, CA 95616, USA; tyson@physics.ucdavis.edu
[2] SpaceX, One Rocket Rd., Hawthorne, CA 90250, USA

## Abstract

The apparent brightness of satellites is calculated as a function of satellite position as seen by a ground-based observer in darkness. Both direct illumination of the satellite by the Sun as well as indirect illumination due to reflection from the Earth are included. The reflecting properties of the satellite components and of the Earth must first be estimated (the Bidirectional Reflectance Distribution Function, or BRDF). The reflecting properties of the satellite components can be found directly using lab measurements or accurately inferred from multiple observations of a satellite at various solar angles. Integrating over all scattering surfaces leads to the angular pattern of flux from the satellite. Finally, the apparent brightness of the satellite as seen by an observer at a given location is calculated as a function of satellite position. We develop an improved model for reflection of light from Earth's surface using aircraft data. We find that indirectly reflected light from Earth's surface contributes significant increases in apparent satellite brightness. This effect is particularly strong during civil twilight. We validate our approach by comparing our calculations to multiple observations of selected Starlink satellites and show significant improvement on previous satellite brightness models. Similar methodology for predicting satellite brightness has already informed mitigation strategies for next-generation Starlink satellites. Measurements of satellite brightness over a variety of solar angles widens the effectiveness of our approach to virtually all satellites. We demonstrate that an empirical model in which reflecting functions of the chassis and the solar panels are fit to observed satellite data performs very well. This work finds application in satellite design and operations, and in planning observatory data acquisition and analysis.

*Unified Astronomy Thesaurus concepts:* Artificial satellites (68)

## 1. Introduction

In recent years, numerous large Low Earth Orbit (LEO) satellite constellations have been proposed. There are currently more than 6000 LEO satellites in operation, a sixfold increase over just two years. This is expected to increase exponentially over the next decade. The impact on astronomy research (Tyson et al. 2020; Hu et al. 2022) and on the night sky environment (Venkatesan et al. 2020; Lawrence et al. 2022; Barentine et al. 2023) has been discussed widely. Technical mitigation involves innovation in satellite design, satellite operations, and astronomy data processing and analysis. The science pursued by ground-based wide-field sky surveys such as Rubin Observatory's Legacy Survey of Space and Time (LSST; Ivezić et al. 2019), as well as all other optical observatories, large and small, is impacted by satellite streaks.

After dusk and before dawn, LEO satellites scatter sunlight onto the Earth's surface. This sunlight is both direct and indirect (reflected from Earth). This scattered light can interfere with both casual stargazing and large ground-based observatories. The net effect depends on several variables including: satellite geometry, satellite material properties, satellite orientation, wavelength, satellite location, observatory location, satellite range, and number of satellites. In order to quantify this effect, it is necessary to predict satellite brightness. To make this prediction, we must measure the material properties of satellite surfaces, either directly in the lab or indirectly using satellite brightness observations. We must also know both the orientations and areas of the satellite's surfaces.

This paper presents techniques for calculating the brightness of satellites seen by observers on the Earth's surface. We consider two sources of light that can be scattered by a satellite. First, there is light directly incident from the Sun. Second, we include light scattered from the portion of Earth's surface illuminated by the Sun and visible to the satellite. We refer to the latter as earthshine. We treat the Sun as a plane-wave source and the Earth as a sphere. Because the geometry and the light sources are relatively simple, we can directly calculate fluxes incident on the satellite. Then, using a simple model for the satellite's reflectance and the position of the satellite in the sky, we calculate the overall satellite brightness. This technique offers a respectable improvement to previous diffuse sphere models of satellites and is computationally efficient. A diagram of the geometry and the light sources is shown in Figure 1.

We model a satellite as a collection of opaque surfaces. The light scattered from each surface is defined by an isotropic Bidirectional Reflectance Distribution Function (BRDF; Greynolds 2015). Physically, this means that a rotation of a surface about its normal vector does not change scattering properties. Even though individual surfaces are isotropic, in most cases the overall effective scatter from the satellite will be anisotropic. The BRDF depends on both the surface's material and surface finish. For example, a smooth metallic surface such as bare aluminum is very specular, but a rough painted surface is mostly diffuse. At this time, we do not consider shadowing between surfaces (mutual shadowing) nor anisotropic BRDFs. If there are complex components on a satellite, a ray-tracing analysis can be used to determine an "effective BRDF" for individual components. Ray tracing allows the inclusion of shadowing and scatter from multiple bounces. We find that this complexity is not required to achieve good observation correlation for Starlink satellite architecture. If lab-measured BRDFs are not available, the BRDF can also be estimated by fitting to satellite brightness observations taken over a variety of solar angles. This corresponds to an indirect measurement of the BRDF.

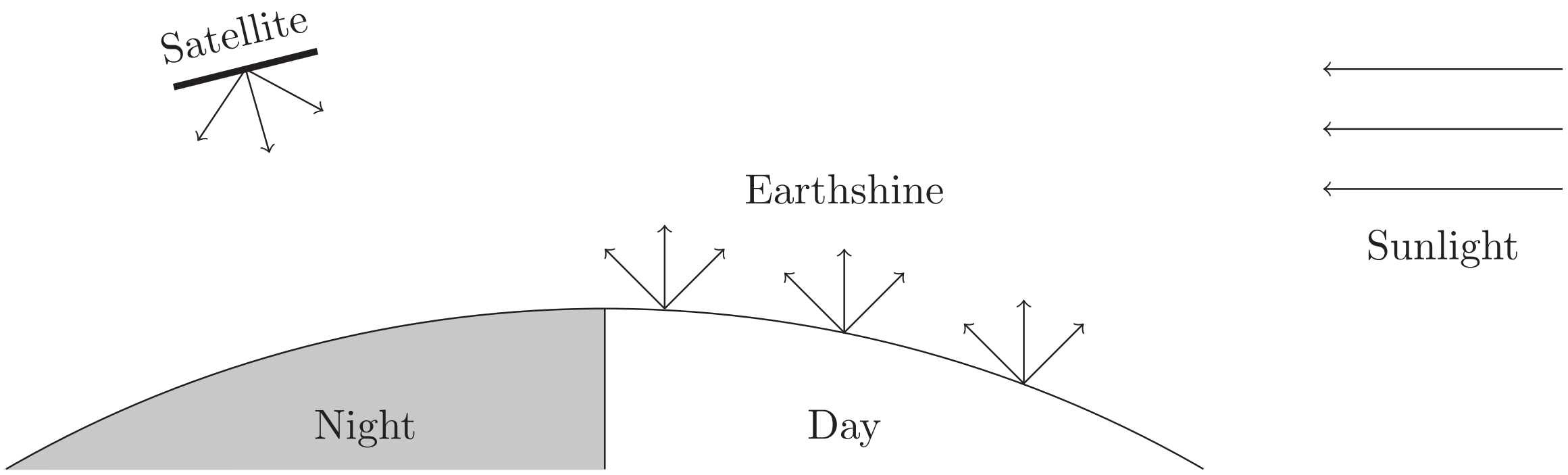

**Figure 1.** Both sunlight and earthshine are scattered by the satellite onto the night side of Earth's surface.

These calculations have been wrapped in a publicly available Python package called `Lumos-Sat` (Fankhauser 2023a). Tools for predicting satellite brightness are critical for both constellation operators and observatories. Constellation operators can use `Lumos-Sat` to include satellite brightness as a design constraint, by quantifying the brightness effects of changing satellite material, geometry, or orientation. Meanwhile, `Lumos-Sat` lets observatories predict and mitigate the impact of existing satellites on science.

## 2. Light Transport

First, we compute a simplified light transport equation. Because the distances between the Sun, the Earth, and the satellite are much larger than the scale of a satellite, we can make a variety of simplifying assumptions. Given a distant light source, a surface, and a distant observer, our goal is to find the flux scattered from the source onto the observer. For a given geometry, the BRDF is a function of the unit vector to the light source $\hat{w}_i$ and the unit vector to the observer $\hat{w}_o$. The BRDF is defined as follows:

$$\mathrm{BRDF} = f_r(\hat{w}_i, \hat{w}_o) \equiv \frac{1}{\cos(\phi_o)\cos(\phi_i)L_i}\frac{\partial L_o}{\partial \hat{w}_i}. \quad (1)$$

The ingoing radiance is $L_i$ and the outgoing radiance is $L_o$. Here, $\phi_i$ is defined as the angle between the surface normal $\hat{n}$ and the vector to the source $\hat{w}_i$. Likewise, $\phi_o$ is defined as the angle between the surface normal $\hat{n}$ and the vector to the observer $\hat{w}_o$.[3] In this analysis, we consider a very distant point source, which can be treated as a plane wave with flux $I_{\mathrm{in}}$ at the surface. The surface has an area $A$. This geometry is shown in Figure 2.

We can therefore rearrange Equation (1) to find:

$$\begin{aligned} L_o &= \int f_r \cos(\phi_i)\cos(\phi_o) L_i dw_i \\ &= \int f_r \cos(\phi_i)\cos(\phi_o) A I_{\mathrm{in}} \delta(w_i) dw_i \\ &= f_r \cos(\phi_i)\cos(\phi_o) A I_{\mathrm{in}}. \end{aligned} \quad (2)$$

The fraction of flux scattered by the surface from the source to an observer at distance $d$ is given:

$$G(\text{source} \to \text{observer}) \equiv \frac{I_{\mathrm{out}}}{I_{\mathrm{in}}} = (\hat{w}_i \cdot \hat{n})(\hat{w}_o \cdot \hat{n}) f_r(\hat{w}_i, \hat{w}_o)\frac{A}{d^2}. \quad (3)$$

Let us recall that $f_r$ is the BRDF of the surface. The fraction of light scattered to the observer increases with surface area perpendicular to the source or observer and decreases with distance as an inverse square law. The angular distribution is determined by the BRDF. We will use the light transport equation given in Equation (3) to calculate both the flux of light scattered by the Earth's surface and the flux scattered by the satellite. The next step in our analysis is to find good BRDF models, so that our light transport equation is accurate.

## 3. BRDF Models

Our brightness calculations will only be as accurate as our BRDF models. We recommend fitting most measured data to a binomial BRDF model. The model parameters are $b_{ik}$, $c_{ik}$, and $q$:

$$\log(\mathrm{BRDF}) = \sum_{k=0}^{n}\left\{\sum_{i=0}^{m} b_{ik}\delta^i + \frac{1}{2}\sum_{i=l_1}^{l_2} c_{ik}\log(1 + q^i\delta^2)\right\}\nu^k, \quad (4)$$

$$\delta^2 = \|\boldsymbol{\rho} - \boldsymbol{\rho}_0\|^2 \qquad \nu = \boldsymbol{\rho}\cdot\boldsymbol{\rho}_0. \quad (5)$$

The vectors $\boldsymbol{\rho}$ and $\boldsymbol{\rho}_0$ are the projection of the outgoing unit vector onto the surface and the projection of the specularly reflected unit vector $\hat{s}$ onto the surface, respectively. Both of these vectors can be written as functions of the incoming vector $\hat{w}_i$, the outgoing vector $\hat{w}_o$, and the surface normal vector $\hat{n}$:

$$\begin{aligned} \boldsymbol{\rho} &= \hat{w}_o - (\hat{w}_o \cdot \hat{n})\hat{n} \\ \boldsymbol{\rho}_0 &= \hat{s} - (\hat{s} \cdot \hat{n})\hat{n} \\ \hat{s} &= 2(\hat{w}_i \cdot \hat{n}) - \hat{w}_i. \end{aligned} \quad (6)$$

We note that $\hat{s}$ is the specularly reflected unit vector.

The binomial model is ideal because it enforces physical realism and has been used extensively and proven in commercial optical analysis (Greynolds 2015). Constructing the BRDF from variables $\delta$ and $\nu$ is convenient because it enforces positivity, continuity, reciprocity, and isotropic scatter. However, is important to note that the binomial model is empirical. The parameters $b_{ik}$, $c_{ik}$, and $g$ must be fit to measured BRDF data. Fitting binomial models does require a degree of caution. The number of coefficients should be kept as small as possible to avoid overfitting. Binomial fits should always be reviewed carefully before using them in a calculation. In general, however, any BRDF model could be used, e.g., the Harvey-Shack model (Nattinger 2020), Ross-Li model (Wanner et al. 1995), or Phong BRDF (Phong 1973). BRDFs can also be interpolated from measured data; however, data are usually taken at a small number of incident angles, so interpolation can lead to extrapolation errors or enhanced measurement noise. Additionally, directly interpolating in spherical coordinates will not work, because the specular peak of the BRDF shifts. For successful interpolation, it is necessary to first make a coordinate transform. We find the best results by interpolating the BRDF as a function of ($\theta_o$, $\phi_o - \phi_s$), where ($\theta_o$, $\phi_o$) is the outgoing direction of scattered light and ($\theta_s$, $\phi_s$) is specularly reflected direction. BRDFs can also be interpolated as a function of $\delta$, as given in Equation (5). When done cautiously, binomial model development will overcome the shortfalls and complexity of interpolation.

[3] When the $\cos(\phi_o)$ term is removed from the denominator of Equation (1), the resulting quantity is referred to as the Cosine Corrected BRDF (CCBRDF) or Angular Resolved Scatter (ARS).

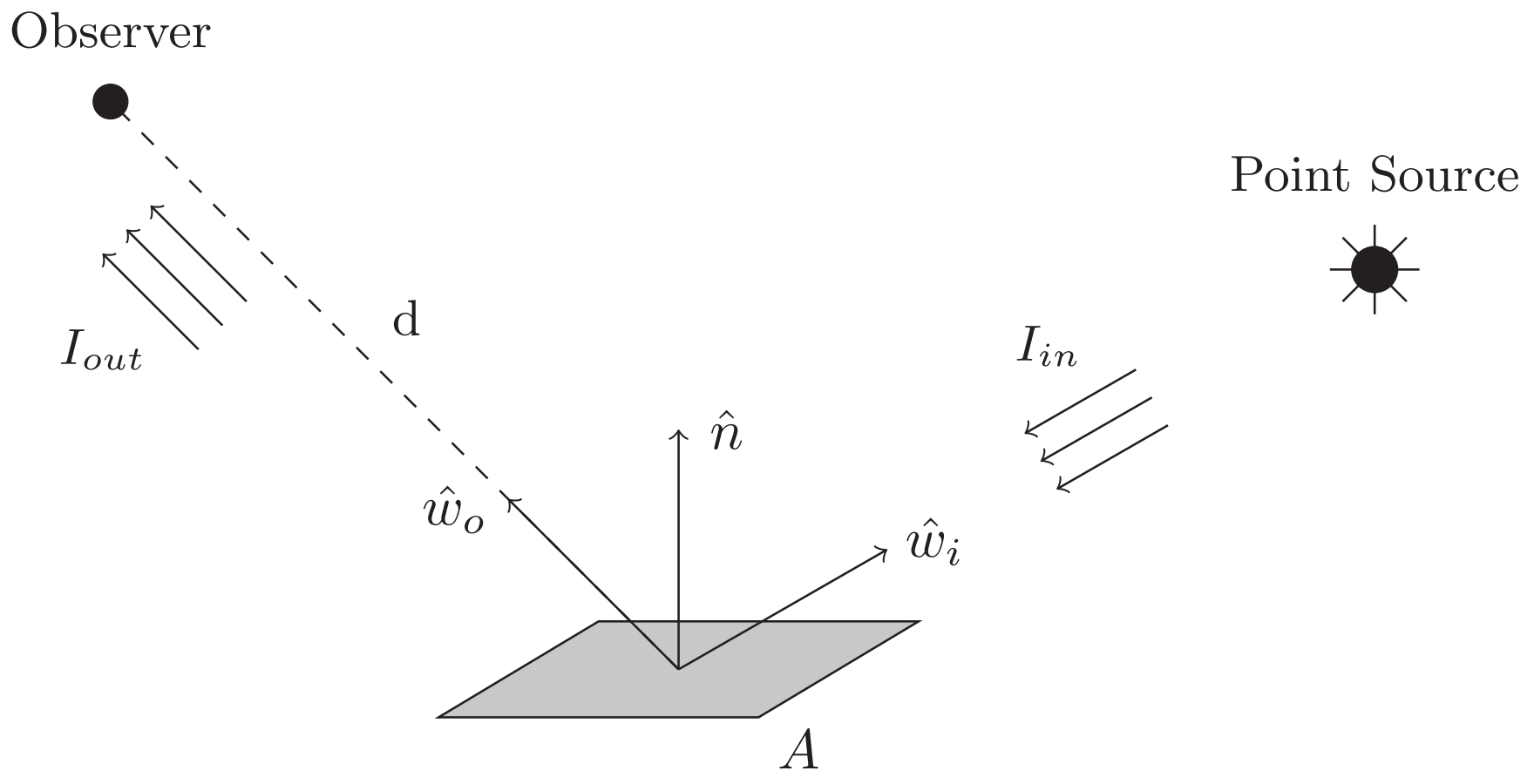

**Figure 2.** The geometry for scattering of light from a point source off of a surface with area A to an observer at range $d$.

The BRDF of each surface on a satellite can be experimentally measured (Germer & Asmail 1997) or estimated from a catalog of known material BRDFs (Matusik et al. 2003). An effective BRDF for Earth's surface can be fit to data gathered from remote sensors. In particular, we use measurements from NASA's Cloud Absorption Radiometer (Gatebe & King 2016). The aircraft's radiometer has one degree angular resolution in 14 spectral bands 340–2300 nm. We use the 479 nm data averaged over hundreds of images, uncorrected for atmospheric absorption. This data is then fit to a Phong BRDF, of the following form:

$$\mathrm{BRDF} = \frac{K_d}{\pi} + K_s\frac{n+2}{2\pi}(\hat{w}_r \cdot \hat{w}_o)^n. \tag{7}$$

The parameter $K_d$ controls the magnitude of the diffuse component of the BRDF, while $K_s$ controls the magnitude of a specular lobe and $n$ controls the width of the specular peak. The vectors $\hat{w}_r$ and $\hat{w}_o$ are the specularly reflected unit vector and the outgoing unit vector, respectively. We find that the Phong model yields more reliable results when fit to aggregate data than does a binomial model.

## 4. Calculations in the Satellite-centered Frame

Our goal now is to use the equation for the fraction of scattered flux $G$(source → observer) given in Equation (3) to calculate the flux scattered by a satellite onto an observer on Earth's surface. We start by introducing a coordinate frame that simplifies this calculation. Next, we find the contribution of light directly scattered by the satellite from the Sun. Finally, we include the contribution of light scattered from earthshine.

### *4.1. The Satellite-centered Frame*

To simplify calculations, we would like to use a reference frame that reduces dependent variables as much as possible. Our choice of frame is called the satellite-centered frame.[4] The $z$-axis points along geodetic zenith. The $y$-axis is in the plane defined by the center of the Earth, the Sun, and the satellite, and is perpendicular to the $z$-axis. The angle between the $y$-axis and the vector from the center of the earth to the Sun must be less than 180°. The $x$-axis is defined by the right-hand rule. This frame is shown in Figure 3.

In the satellite-centered frame, the flux seen by an observer depends only on the angle of the satellite past terminator $\alpha$, the vector from the satellite to the observer $\boldsymbol{v}$, the radius of Earth $R_{\mathrm{E}}$, and the geodetic height of the satellite $h$.

### *4.2. Flux from the Sun*

The fraction of flux scattered by a single surface has been derived in Equation (3). We simply sum over all $N_s$ surfaces in a satellite to arrive at:

$$I_{\mathrm{observer}} = I_{\mathrm{sun}}\sum_{s=1}^{N_s} G_s(\mathrm{sun} \rightarrow \mathrm{observer})$$
$$G_s(\mathrm{sun} \rightarrow \mathrm{observer}) = \frac{A_s f_s(\hat{v}_{\mathrm{sat}\rightarrow\mathrm{sun}}, \hat{v}_{\mathrm{sat}\rightarrow\mathrm{obs}})\mathcal{N}_s}{\|\boldsymbol{x}_{\mathrm{obs}} - \boldsymbol{x}_{\mathrm{sat}}\|^2}$$
$$\mathcal{N}_s = (\hat{n}_s \cdot \hat{v}_{\mathrm{sat}\rightarrow\mathrm{sun}})(\hat{n}_s \cdot \hat{v}_{\mathrm{sat}\rightarrow\mathrm{obs}}). \tag{8}$$

[4] The satellite-centered frame should not be confused with a satellite body frame, which is fixed to the satellite's chassis and has an origin at the satellite's center of mass.

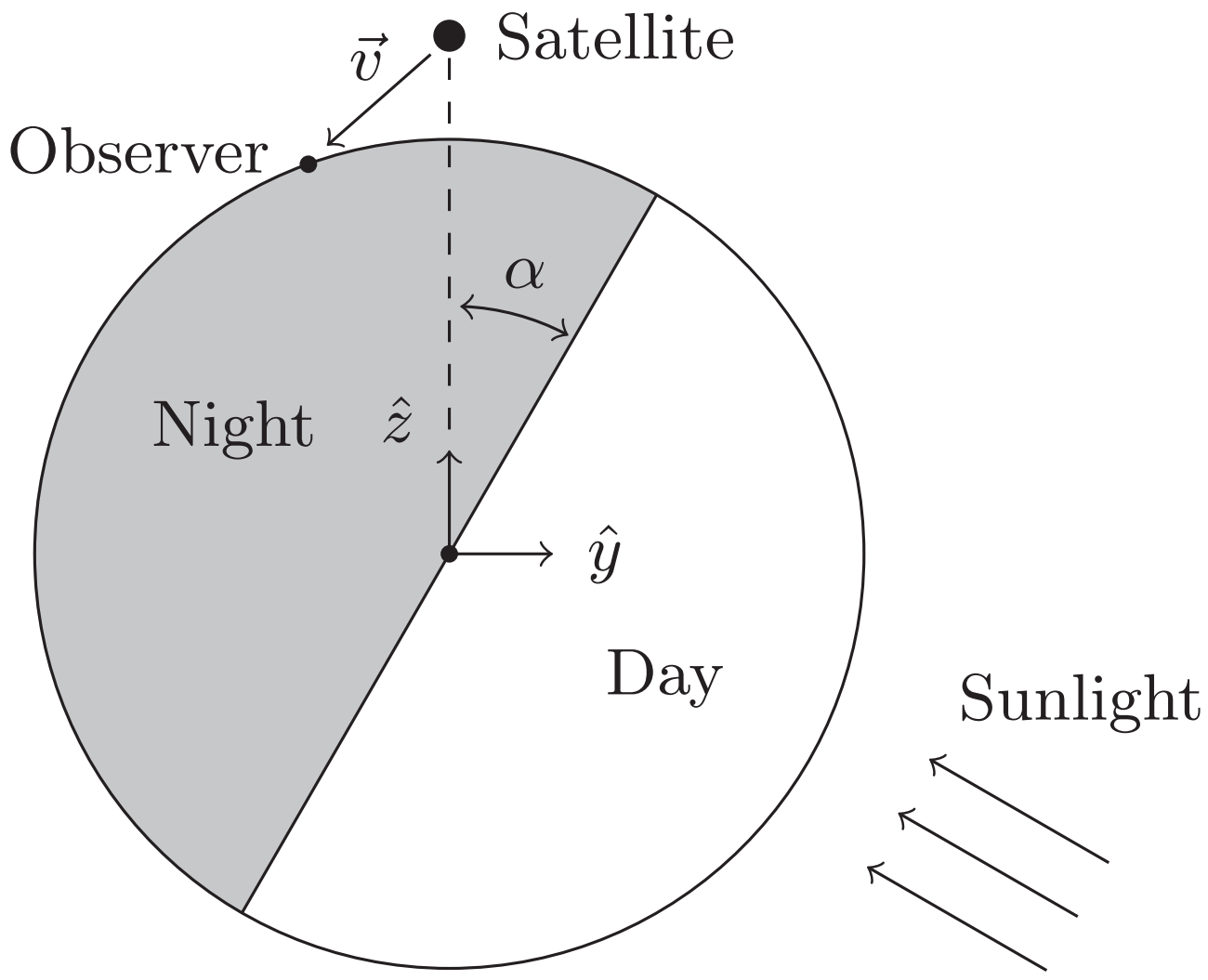

**Figure 3.** The geometry of the satellite-centered frame. $\hat{x}$ is out of the page.

The area, BRDF, and normal of satellite surface $s$ are $A_s$, $f_s$, and $\hat{n}_s$, respectively. The distance from the satellite to the observer is $\|\boldsymbol{x}_{\text{obs}} - \boldsymbol{x}_{\text{sat}}\|$. The unit vector from the satellite to the Sun is $\hat{v}_{\text{sat}\to\text{sun}}$ and the unit vector from the satellite to the observer is $\hat{v}_{\text{sat}\to\text{obs}}$.

### 4.3. Flux from Earthshine

The flux seen by the observer caused by earthshine can be calculated similarly. Light is scattered first by the Earth's surface, then by the satellite. The flux scattered by a single satellite surface is calculated by integrating over the portion of Earth's surface $E$ that is illuminated by the Sun and is visible to the satellite. This is then summed over the number of satellite surfaces $N_s$. We find the flux seen by an observer due to earthshine:

$$I_{\text{observer}} = I_{\text{sun}} \sum_{s=0}^{N_s} \iint_E G_s(\partial A \to \text{obs}) \cdot \partial G(\text{sun} \to \text{sat}). \tag{9}$$

$\partial A$ is a differential area of Earth's surface. $G_s$ is the fraction of earthshine flux scattered from $\partial A$ to the observer by a surface. $\partial G$ is the differential fraction of Sun flux scattered from the Sun to the satellite by $\partial A$. We can approximate Equation (9) by discretizing the integral. This yields:

$$I_{\text{observer}} = I_{\text{sun}} \sum_{s=0}^{N_s} \sum_{p=0}^{N_p} G_s(\Delta A_p \to \text{obs}) \cdot \Delta G_p(\text{sun} \to \text{sat}). \tag{10}$$

We refer to each area element on the Earth's surface $\Delta A_p$ as a patch. As the number of patches on the Earth's visible and illuminated surface, $N_p$, goes to infinity, Equations (9) and (10) become equivalent. The fraction of earthshine scattered to the observer is then given by

$$G_s(\Delta A_p \to \text{obs}) = \frac{A_s f_s(\hat{v}_{\text{sat}\to\text{patch}}, \hat{v}_{\text{sat}\to\text{obs}})\mathcal{N}_s}{\|\boldsymbol{x}_{\text{obs}} - \boldsymbol{x}_{\text{sat}}\|^2}$$
$$\mathcal{N}_s = (\hat{n}_s \cdot \hat{v}_{\text{sat}\to\text{patch}})(\hat{n}_s \cdot \hat{v}_{\text{sat}\to\text{obs}}). \tag{11}$$

Let us recall again that the area, BRDF, and normal of satellite surface $s$ are $A_s$, $f_s$, and $\hat{n}_s$, respectively. The distance from the satellite to the observer is $\|\boldsymbol{x}_{\text{obs}} - \boldsymbol{x}_{\text{sat}}\|$. The unit vector from the satellite to the patch on the Earth's surface is $\hat{v}_{\text{sat}\to\text{patch}}$, and the unit vector from the satellite to the observer is $\hat{v}_{\text{sat}\to\text{obs}}$.

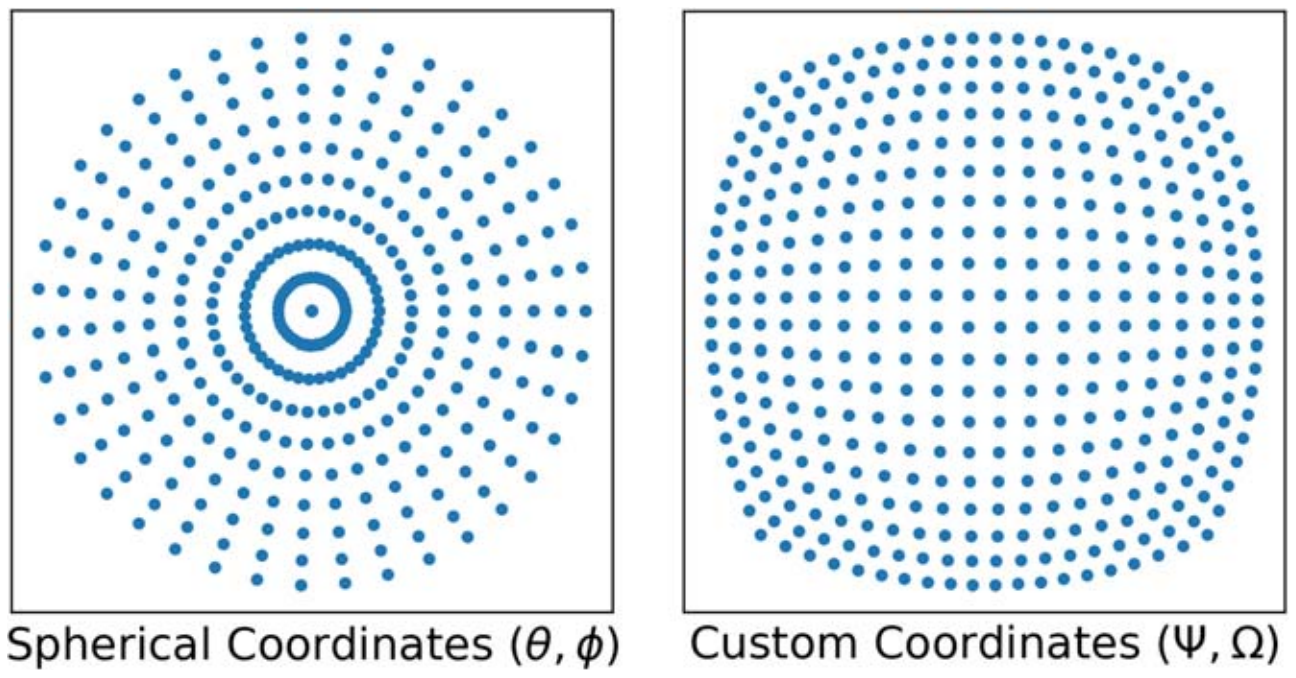

**Figure 4.** We compare discretizations schemes by uniformly sampling 400 patches in both standard spherical coordinates and our custom coordinates. We see that our custom coordinate system generates a discretization that is more evenly distributed over the Earth's surface.

The fraction of Sun flux scattered from the Sun to the satellite by a differential area of of Earth's surface is

$$\Delta G_p(\text{sun} \to \text{sat}) = \frac{\Delta A_p f_p(\hat{v}_{\text{sun}\to\text{patch}}, \hat{v}_{\text{patch}\to\text{sat}})\mathcal{N}_p}{\|\boldsymbol{x}_{\text{sat}} - \boldsymbol{x}_{\text{patch}}\|^2}$$
$$\mathcal{N}_p = (\hat{n}_p \cdot \hat{v}_{\text{patch}\to\text{sun}})(\hat{n}_p \cdot \hat{v}_{\text{patch}\to\text{sat}}). \tag{12}$$

The area, BRDF, and normal of a patch $p$ on the Earth's surface are $\Delta A_p$, $f_p$, and $\hat{n}_p$, respectively. The distance from the satellite to the observer is $\|\boldsymbol{x}_{\text{obs}} - \boldsymbol{x}_{\text{sat}}\|$. The unit vector from the satellite to the patch on the Earth's surface is $\hat{v}_{\text{sat}\to\text{patch}}$, and the unit vector from the satellite to the observer is $\hat{v}_{\text{sat}\to\text{obs}}$.

### 4.4. Discretization of Earth's Surface

It is now necessary to discretize the portion of Earth that is visible to the satellite and illuminated by the Sun. At first glance, using standard spherical coordinates seems like the easiest solution. Unfortunately, this results in a discretization that is heavily weighted at the poles of the Earth. We instead propose the following coordinate system:

$$x = z \tan\psi$$
$$y = z \tan\Omega$$
$$z = \frac{R_{\text{E}}}{\sqrt{1 + \tan^2\psi + \tan^2\Omega}}. \tag{13}$$

The variables $\psi$ and $\Omega$ represent the angle off-plane and angle on-plane, respectively, where the plane is defined by $\hat{y}$ and $\hat{z}$. The radius of the Earth is given as $R_{\text{E}}$.

Using this coordinate system results in patches that have much more even spacing. Figure 4 shows 400 patches covering a quarter-sphere. We can see that using standard spherical coordinates results in "bunching up" at the poles.

In order to calculate the flux, we need to know the area $\Delta A_p$ of each patch $p$ in the mesh. This is approximated using the Jacobian determinant as follows:

$$\Delta A_p = \frac{\partial(x, y, z)}{\partial(\psi, \Omega, R_{\text{E}})} \Delta\psi\Delta\Omega. \tag{14}$$

Finally, we must only include patches that are both visible to the satellite and illuminated by the Sun. Consider a patch at the point $(x, y, z)$, measured in the satellite-centered frame. The satellite is visible to a patch if the following is true:

$$\cos\left(\frac{R_{\rm E}}{R_{\rm E}+h}\right) < \frac{z}{R_{\rm E}}. \tag{15}$$

A patch is illuminated by the Sun if $y > 0$. Using Equation (13), these two constraints can be related back to the angle off-plane and the angle on-plane, $(\Psi, \Omega)$.

In practice, we find that discretization causes some noise in our brightness calculations. The amplitude of the noise decreases with number of patches and the frequency increases. We recommend applying some smoothing to results for calculations which include earthshine.

### 4.5. Converting Flux to AB Magnitude

It is useful to convert from incident flux to AB magnitude. This allows for comparison between satellite brightness and the brightness of celestial objects. This conversion is simply:

$$\text{ABmagnitude} = -2.5\log_{10}\left(\frac{I}{f}\right) - 56.1. \tag{16}$$

The flux incident on the observer from the satellite is $I$ in units of $\mathrm{W/m^2}$, and $f$ is the frequency of the light in Hz. AB magnitude is defined for a flat spectrum, such as a very hot star. The apparent AB magnitude in a given spectral filter can be found by integrating the flux over the filter's bandpass.

### 4.6. Other Light Sources

Although it is beyond the scope of this paper to include in our simulations, we would also like to offer a "back-of-the-envelope" calculation for the brightness contributions of other light sources. These potentially include celestial objects such as stars, planets, and the Moon. We can modify Equation (3) to estimate the flux scattered by a surface from a source to an observer:

$$I_{\rm observer} = I_{\rm source}\frac{\cos\phi_i\cdot\cos\phi_o\cdot \text{BRDF}\cdot A}{d^2}. \tag{17}$$

Let us recall that $\phi_i$ is the angle between the surface normal and the source and $\phi_o$ the angle between the surface normal and the observer, $A$ is the area of the surface, and $d$ is the range of the satellite. To find an order-of-magnitude calculation for the worst-case scenario, $\cos\phi_i \approx \cos\phi_o \approx 1$, $A \approx 1m^2$, and $d \approx 250km$ (very low Earth orbit). We assume light from our celestial source is very specularly reflected, so that $BRDF \approx 10^4$. We can then use Equation (16) to convert Equation (17) to AB magnitude and find:

$$(\text{ABmag})_{\rm observer} \sim (\text{ABmag})_{\rm source} + 17. \tag{18}$$

From this, we see that, in a worst-case scenario, light scattered from a celestial source onto an observer will be 17 AB magnitude dimmer than the incident light from the source. This means that light from the full Moon could potentially cause brightness up to 4 AB magnitude and Venus up to 12 AB magnitude. The brightest stars like Vega could cause satellite brightness of only around 17 AB magnitude. Using data from NASA's Visible Infrared Imaging Radiometer Suite (VIIRS; Elvidge et al. 2017), we estimate that city lights cause brightness of roughly 15 AB magnitude. Because LEO satellites are not point sources and they move quickly across the focal plane of a telescope, it is unlikely that light scattered from stars or city lights will be detectable by observatories like LSST.

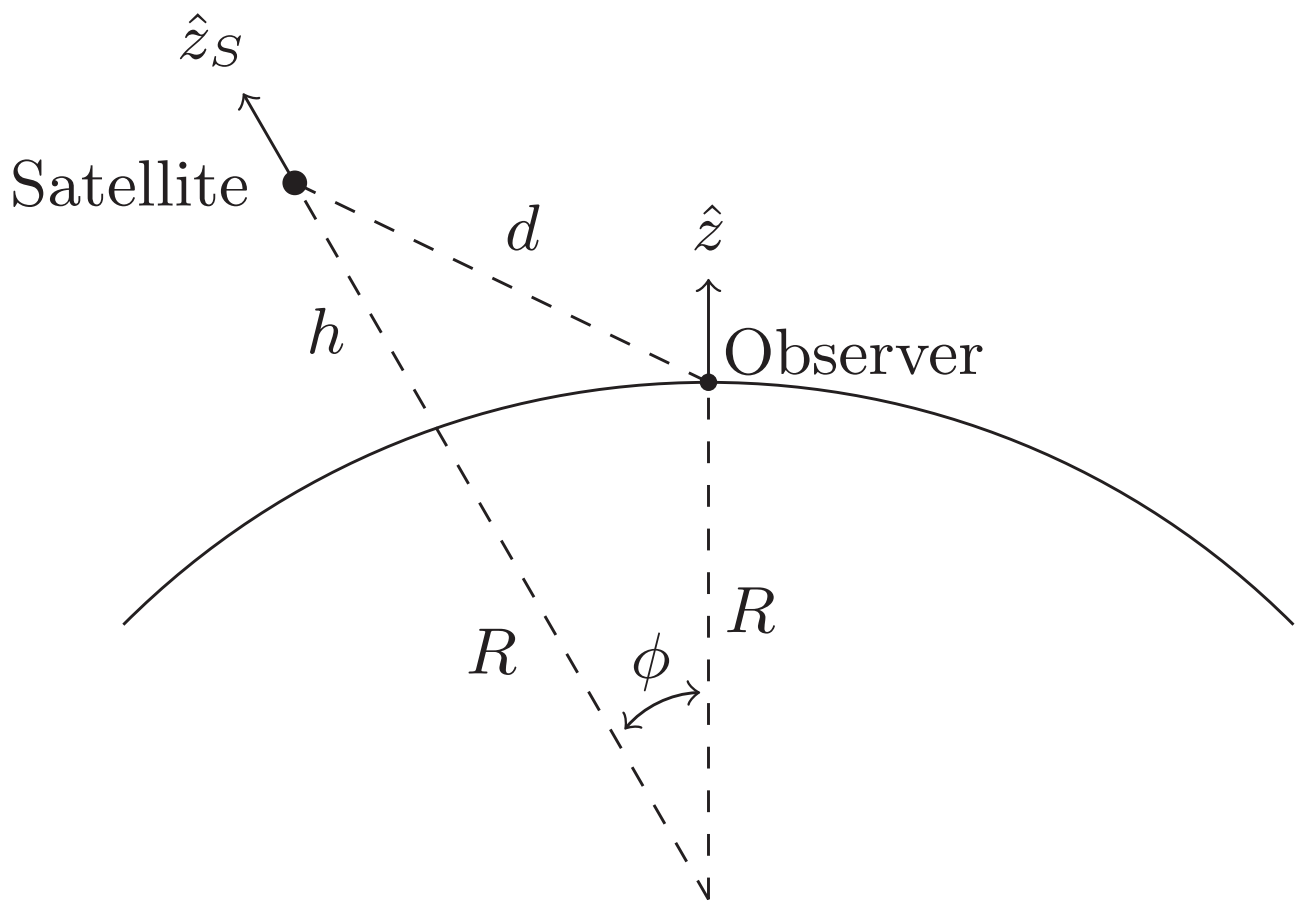

**Figure 5.** Geometry as seen in the observer's frame $O$.

For a typical LEO satellite at 500 km, the motion across the focal plane is about 0°.5 per second. This means the effective exposure time on a 0″.7 PSF footprint is just 0.5 milliseconds, independent of the camera exposure time. For LSST, a 0″.7 PSF has 50 pixels. This means the peak LEO satellite trail surface brightness of a 15 AB magnitude satellite is only 1.8 electrons per pixel. This is buried in the night sky noise from a 15 second LSST exposure.

It is outside the scope of this paper, but future work in satellite brightness modeling should seek to incorporate incident light from the Moon and possibly Venus.

## 5. Calculations in the Observer Frame

The satellite-centered frame is ideal for calculations, but we need to know what brightness a ground-based observer will see. We use the horizontal coordinate system shown in Figure 5. We note that $\hat{z}$ corresponds to an altitude angle of 90°. Given the position of an observer on Earth and the position of a satellite in the sky, our goal is to transform variables from the satellite-centered frame $S$ to the observer frame $O$.[5] We are given a unit vector from the observer to the satellite $[\hat{v}_{\rm sat}]_O$ and the satellite's geodetic height $h$. The unit vector from the observer to the satellite as well as the satellite's height can either be measured by an observer or calculated using the satellite's orbital elements. We also know the unit vector toward the Sun $[\hat{v}_{\rm sun}]_O$, which can be calculated using the latitude and longitude of the observer and the time of observation. The basis vectors in the observer's frame are

$$\begin{aligned}[\hat{x}]_O &= (1, 0, 0)\\ [\hat{y}]_O &= (0, 1, 0)\\ [\hat{z}]_O &= (0, 0, 1).\end{aligned} \tag{19}$$

[5] A similar process to the one shown below can be used to convert to or from the satellite-centered frame. In particular, satellite operators can use these transformations to convert surface normal vectors from a satellite body frame to our satellite-centered frame. This allows brightness calculations to be generated in simulations or in real time using data from onboard satellite sensors.

Moving forward, we will drop the basis notation for vectors in the observer's frame, O. Quantities that are measured in the satellite-centered frame will be marked with an $S$.

We first need to find the basis vectors of the satellite-centered frame, as measured in the observer's frame.

From geometric inspection, we find the following:

$$\phi = \arccos(\hat{z} \cdot \hat{v}_{\rm sat}) - \arcsin\left(\|\hat{v}_{\rm sat} \times \hat{z}\| \cdot \frac{R_{\rm E}}{R_{\rm E} + h}\right), \quad (20)$$

$$\theta = \arctan 2(\hat{y} \cdot \hat{v}_{\rm sat}, \hat{x} \cdot \hat{v}_{\rm sat}), \quad (21)$$

$$d^2 = R_{\rm E}^2 + (R_{\rm E} + h)^2 - 2R_{\rm E}(R_{\rm E} + h)\cos\phi. \quad (22)$$

Let us remember that all vectors are in the observer's frame. The range of the satellite is $d$. The angle $\phi$ is the angular separation between the $z$-axis of the satellite-centered frame and the $z$-axis of the observer's frame. Angle $\theta$ is the rotation of the satellite-centered frame's $z$-axis as measured from the $x$-axis of the observer frame. Using these values, we can find an expression $[\hat{z}_S]_O$, which is the $\hat{z}$ of the satellite-centered basis measured in the observer frame:

$$[\hat{z}_S]_O = (\sin\phi\cos\theta)\hat{x} + (\sin\phi\sin\theta)\hat{y} + (\cos\phi)\hat{z}. \quad (23)$$

It is defined that $\hat{y}_S$ is in the plane containing the vector toward the Sun and $\hat{z}_S$. Additionally, $\hat{y}_S$ is orthonormal to $\hat{z}_S$. This gives three constraints which fully define $\hat{y}_S$:

$$\hat{y}_S = a\hat{z}_S + b\hat{v}_{\rm sun}\hat{y}_S \cdot \hat{z}_S = 0 \qquad \|\hat{y}_S\| = 1. \quad (24)$$

We can solve the three constraints given in Equation (24) to find $a$ and $b$. Physically, $a$ and $b$ are the components of $\hat{y}_S$ in the $\hat{z}_S$ direction and the $\hat{v}_{\rm sun}$ direction, respectively:

$$b = \frac{1}{\sqrt{1 - (\hat{z}_S \cdot \hat{v}_{\rm sun})^2}} a = -b(\hat{z}_S \cdot \hat{v}_{\rm sun}). \quad (25)$$

Finally, $\hat{x}_S$ is defined by the right-hand rule:

$$\hat{x}_S = \hat{y}_S \times \hat{z}_S. \quad (26)$$

The transform from the observer reference frame to the satellite-centered frame is

$$T = T_{\rm observer\rightarrow satellite} = T^{-1}_{\rm satellite\rightarrow observer} = [\hat{x}_S, \hat{y}_S, \hat{z}_S]^{-1}. \quad (27)$$

We can then find the quantities we need to do calculations in the satellite-centered frame. The vector from the satellite to the observer in the satellite-centered frame is

$$[\hat{v}_{\rm sat\rightarrow obs}]_S = (R_{\rm E} + h)\hat{z} - d(T \cdot [\hat{v}_{\rm sat}]_O). \quad (28)$$

Second, the angle of the satellite past the terminator is given:

$$\alpha = -\arcsin(\hat{z} \cdot T \cdot [\hat{v}_{\rm sun}]_O). \quad (29)$$

Let us recall that $R_{\rm E}$ is the radius of the earth, $h$ is the geodetic height of the satellite, and $d$ is the satellite's range. Here, $[\hat{v}_{\rm sat}]_O$ is the vector from the observer to the satellite as measured in the observer's frame, and $[\hat{v}_{\rm sun}]_O$ is the vector from the observer to the Sun as measured in the observer's frame.

If we are interested in a particular satellite, we must know that satellite's position in the sky relative to an observer and geodetic height. These quantities can be found using a satellite's orbital elements. SpaceX and other constellation operators publish traditional TLEs (Two-Line Element Sets) on Space-Track.org. For Starlink, more accurate supplemental TLEs are published on celestrak.org. These supplemental TLEs are fit to Starlink propagated ephemerides and covariances, which are available from "Public Files" on Space-Track.org.

Once a satellite's orbital parameters are known from a TLE, its position at a past or future time can be calculated using a Simplified General Perturbations algorithm. In particular, we use the SGP4 Python package. A satellite's altitude and azimuth as seen from a given location on Earth is then found using tools provided by the Astropy software (Astropy Collaboration et al. 2022).

## 6. Model Validation

In order to validate our calculations, we create a simple brightness model for existing satellite-selected configurations of Starlink V1.5 and compare our AB magnitude calculations to observations. We show that the BRDFs for each satellite surface can either be measured in a laboratory or found by fitting to brightness measurements. We also compare our calculations and observations to the previously standard satellite brightness model—a diffuse sphere.

### *6.1. Starlink v1.5 Brightness Model*

The two largest surfaces on Starlink v1.5 are the solar array and the chassis. The solar array has an area of 22.68 $m^2$, and the chassis nadir has an area of 3.64 $m^2$. SpaceX has provided BRDF data for each surface and information about the normal vectors of these surfaces in the brightness regime of interest. Using this information, we can create a brightness model for a subset of the Starlink v1.5 satellites. The Starlink v1.5 satellites have gone through a variety of design changes to reduce brightness, so we only consider the Starlink v1.5 satellites with the latest brightness configuration. These satellites have a reflective sticker on the chassis nadir and dark pigmented backsheet on the solar arrays (SpaceX 2022). In nominal operations, the chassis points directly nadir and the solar array is perpendicular to the chassis and in the direction of the Sun.

SpaceX has also provided experimentally measured BRDF data for each of these surfaces. This SpaceX contracted BRDF data was measured by Scatterworks using an SS4 scatterometer operating at a wavelength of 532 nanometers. The data is taken at an angular resolution of 1°. This BRDF data is fit to binomial models, using the methods described in Section 3. The data and the fits are shown in Figure 6.

To find a representative BRDF for Earth's surface, we fit Phong models to CARs data, a technique also described in Section 3. We use data from two missions. First, the CLASIC mission, which gathered BRDF data for vegetation over Oklahoma. Using this data to construct an "effective BRDF" for generic vegetation, we find parameters $K_d = 0.53$, $K_s = 0.28$, and $n = 7.31$. Second, the CLAMS mission, which gathered BRDF data for the ocean off the East Coast of the United States. For ocean water, we find $K_d = 0.48$, $K_s = 0.08$, and $n = 16.45$.

We can then feed our knowledge of the satellite's primary surfaces and BRDF models for the Earth's surface into our software and calculate satellite brightness.

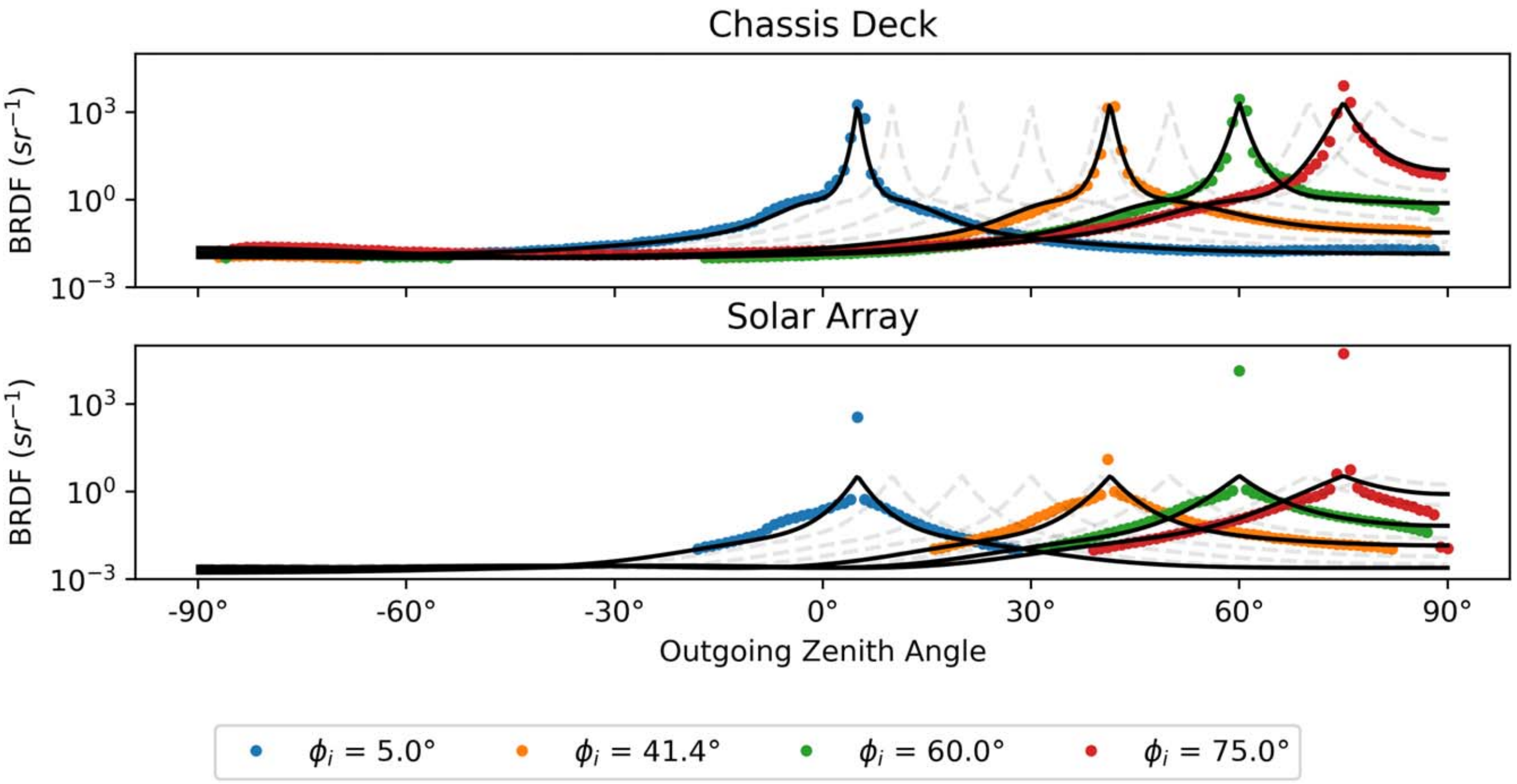

**Figure 6.** Measured data and BRDF fits for the primary surfaces of Starlink v1.5 Satellites. Data are shown at four incident angles. The ingoing and outgoing directions are given in spherical coordinates ($\theta_i$, $\phi_i$) and ($\theta_o$, $\phi_o$), respectively, where $\theta$ is the azimuthal angle and $\phi$ is the zenith angle. All BRDF data are "in-plane", so $\theta_i = 180°$ and $\theta_o = 0°$. Dashed gray lines show the BRDF at intermediate angles, spaced 10° apart. The solar array fit misses the specular peak, but because there is no orientation where light is specularly reflected by the solar array onto observers, this is not a problem.

### 6.2. Starlink v1.5 Model without BRDF Data

For many satellites, BRDF measurements for primary surfaces may not be available. Not only do BRDF measurements require specialized equipment, but also collaboration between some satellite operators and astronomers may prove difficult. In this case, we need to use a different approach to find an accurate satellite brightness model. We assume minimal information of the satellite—just the pointing directions of primary surfaces. For Starlink v1.5, these surfaces are the chassis, which points directly nadir, and the solar array, which is perpendicular to the chassis and toward the Sun. We then assign each of these two surfaces a Phong BRDF, which has three free parameters. The areas of each surface are set to 1 $m^2$—this means that the best-fit albedo of each surface may be greater than unity. In total, our empirical satellite model has six unknown parameters. These parameters can be found by fitting to observed Starlink brightness data. As long as there are a variety of observations over many different solar angles, this makes for an accurate brightness model. This technique uses brightness observations to indirectly measure the overall effective BRDF of the satellite. The satellite's BRDF is then decomposed into the individual surface BRDFs by fitting to brightness observations. The ability to find effective BRDFs for satellite surfaces from only ground-based brightness data is extremely important to astronomers. It makes our brightness modeling approach viable for most satellites.

### 6.3. Diffuse Sphere Model

We can also compare our brightness model to the commonly used diffuse sphere model. For this model, the flux of light scattered by the satellite and incident on an observer is simply:

$$I = \frac{I_{\rm sun}}{d^2}\frac{2}{3\pi^2}A\rho((\pi - \phi)\cos\phi + \sin\phi). \quad (30)$$

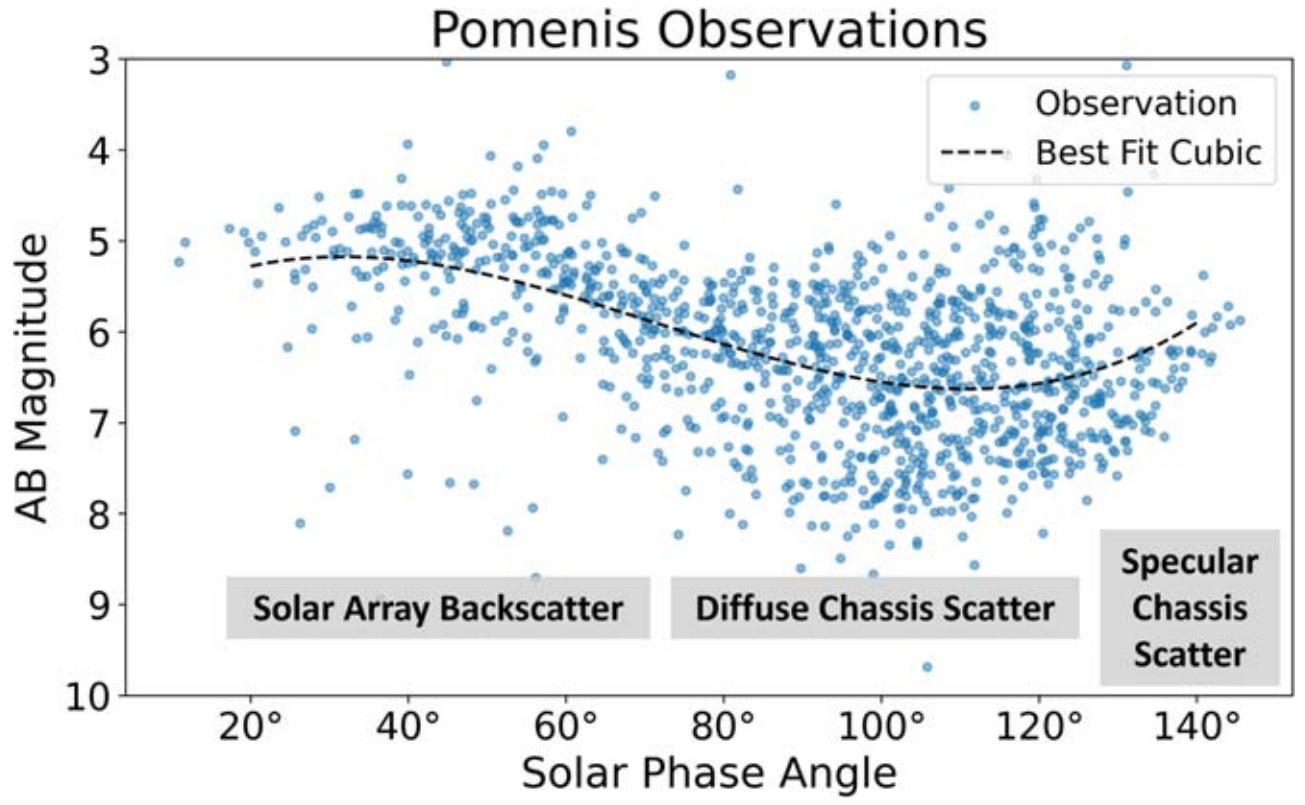

**Figure 7.** Plot of measured Starlink v1.5 brightness vs. solar phase angle. At low solar phase angles, brightness is dominated by backscatter from the solar array. At intermediate phase angles, light is diffusely scattered by the chassis. At high phase angles, light is more specularly reflected by the chassis. We note that second-generation Starlinks will off-point solar arrays and use a chassis material with lower diffuse reflection. These two changes reduce brightness at low and intermediate phase angles.

Here, $\phi$ is the solar phase angle—the angle between the observer, the satellite, and the Sun. The effective area and effective albedo of the satellite are $A$ and $\rho$, respectively. The satellite's range is $d$. The flux of the Sun incident on the satellite is $I_{\rm Sun}$. There is one free parameter, the albedo-area product $\rho A$, which must be best fit to satellite brightness observations. While this model is extremely simple, it has little basis in reality and does not correlate well with brightness measurements. As shown in Figure 7, Starlink v1.5 brightness has a complicated dependence on solar phase angle. This dependence is not fully captured by the diffuse sphere model.

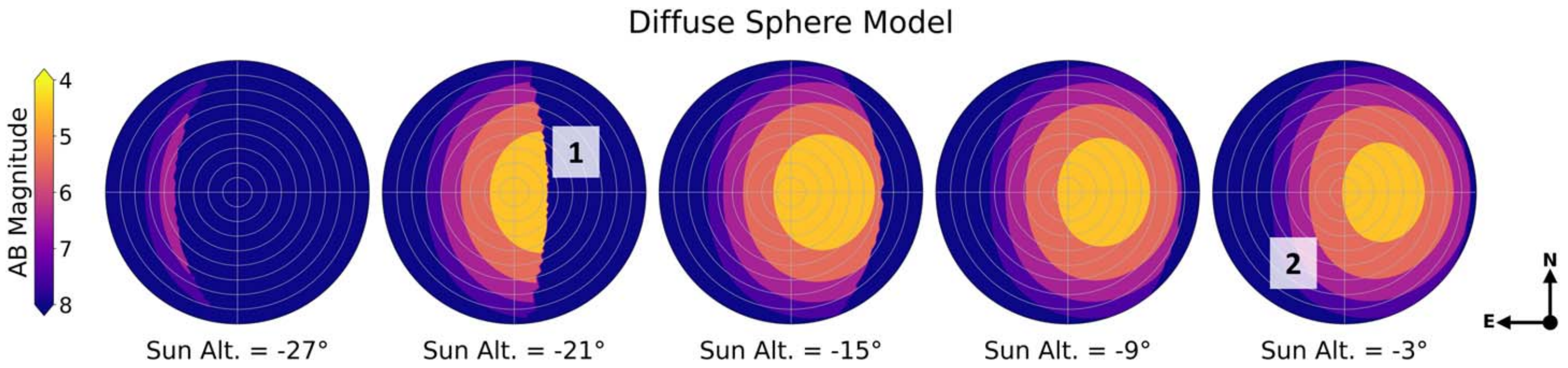

**Figure 8.** The diffuse sphere model shows little variation in brightness. Satellites in the western sky are shadowed by Earth (1). The controlling factors of brightness are satellite range and the illuminated fraction of the sphere. Satellite brightness falls off with satellite range (2). Modeling satellite brightness as a function of just these two variables is overly simplistic.

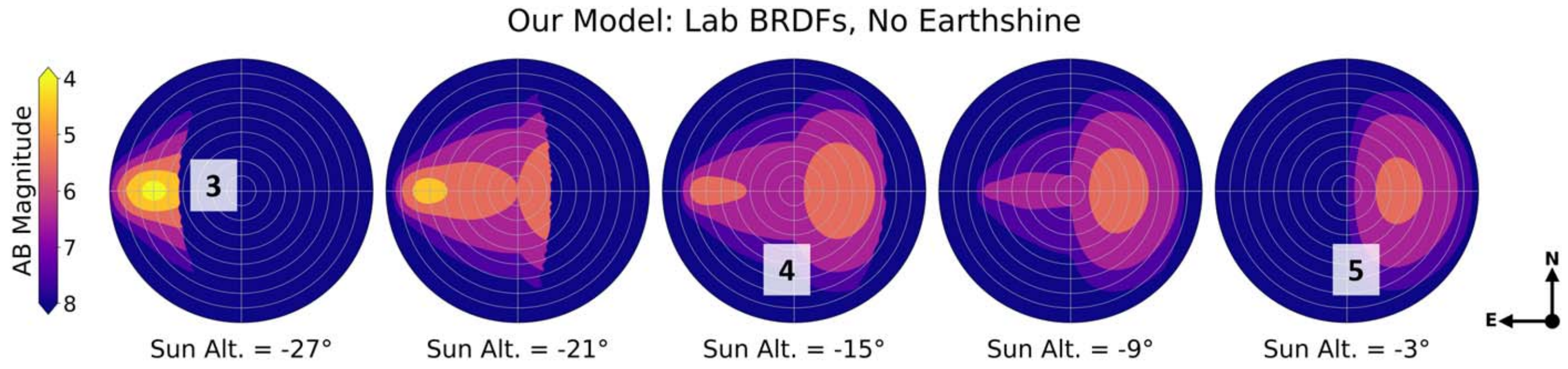

**Figure 9.** In our model, two distinct brightness peaks are clearly visible. The first is visible earlier in the morning and is caused by light forward-scattered from the chassis nadir. This specular peak is visible just above the eastern horizon (3). The second brightness peak is caused by light backscattering from the solar array. It is most prominent in the western sky near dawn (5). There is also a transition where both peaks are visible (4).

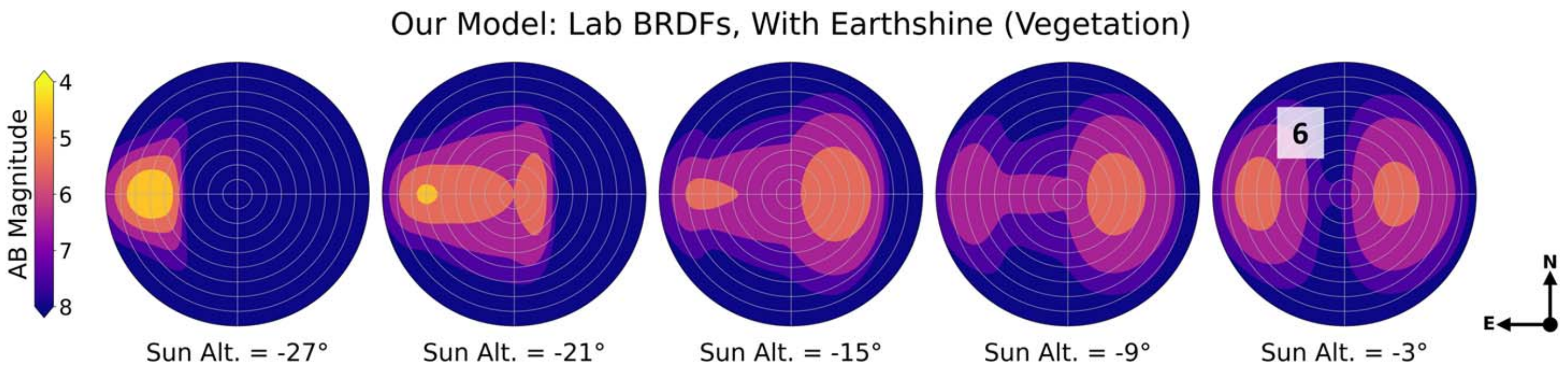

**Figure 10.** Earthshine from vegetation adds an additional component of brightness, low in the eastern sky (6). This effect becomes more and more pronounced as the Sun rises.

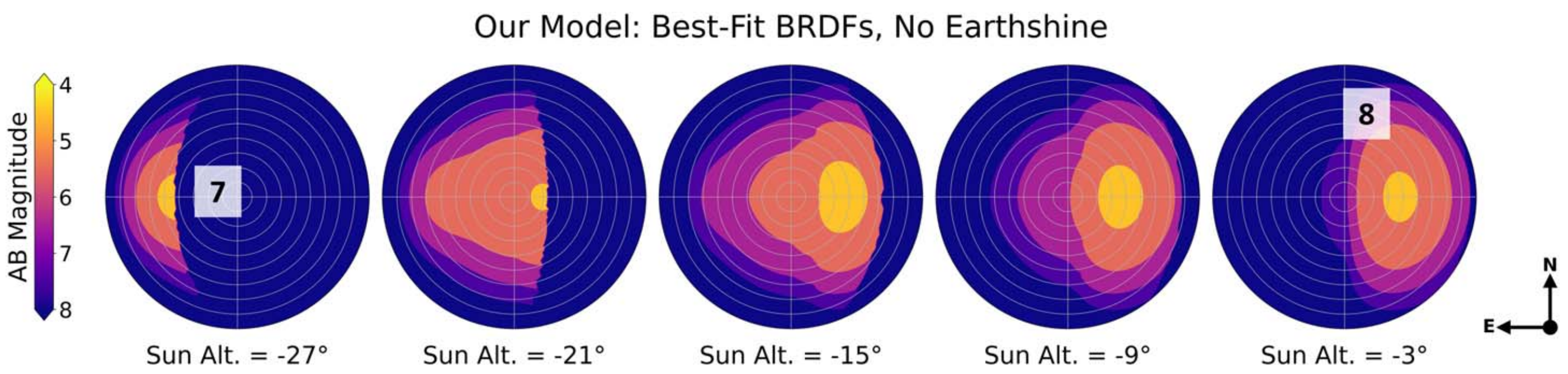

**Figure 11.** With BRDFs best fit to observed data, our model captures both forward scatter from the chassis (7) and backscatter from the solar array (8) to some degree. This model could be improved by using more Starlink v1.5 observations over a wider variety of solar angles and less noisy observations.

### 6.4. Brightness Distributions on the Night Sky

We run our calculations for Sun altitudes ranging from $-27°$ to $-3°$. The Sun's azimuth is fixed at 90°. This corresponds to a period of time prior to sunrise for an observer on the Equator. Satellite altitude is fixed at 550 km[6]. We use our new brightness models both with and without the earthshine contributions. For calculations with earthshine, we use an earthshine discretization of $151 \times 151$ patches. The diffuse sphere calculation is also shown, to provide a comparison. Results are shown in Figures 8–11. It should be noted that these plots show the brightness that a Starlink v1.5 satellite would appear if it was at a given point on the sky. The center of each image corresponds to a satellite directly zenith. The outer edge is the horizon. Cardinal directions are marked: north is toward the top of the image and east is toward the left side. The grid lines show altitude increments of 10° and azimuth increments of 90°.

Our brightness modeling shows significantly different patterns over the night sky compared to the diffuse sphere model. The diffuse sphere model shows little change in brightness with respect to Sun altitude or satellite position. Additionally, the diffuse model predicts that the brightest satellites will be in the western sky, throughout dawn. On the other hand, our models shows how the location of peak satellite brightness changes from the eastern to the western sky as the Sun rises. Earlier in the morning, bright satellites in the eastern sky are caused by light forward scattering off of the specular chassis. Near dawn, bright satellites in the western sky are caused by light backscattered by the solar array. We see that earthshine creates additional brightness in the eastern sky, particularly during civil and nautical twilight. Earthshine drives satellite brightness up to AB magnitude in a region of the sky where satellites are predicted to be invisible when only illuminated directly by the Sun.

### 6.5. Comparison to Observed Brightness

Using SpaceX contracted data gathered at Mount Lemmon in Arizona by the Pomenis Observatory (Pearce et al. 2018), we can compare our brightness calculations to actual observations of satellite brightness. A scatter plot of these observations is shown in Figure 7. For this correlation, we once again only use observations of Starlink satellites with a pigmented solar array backsheet and reflective chassis sticker. We also exclude observations of satellites that do not have their solar array and chassis in the orientation we used for brightness modeling. The majority of these excluded observations are from satellites, which were raising or lowering their orbit at the time of observation. Starlink satellites performing orbital maneuvers use a distinct brightness mitigation orientation as much as possible. Despite removing these observations, the data set is still somewhat noisy. The Pomenis observatory has a typical measurement error of 0.1 AB magnitude. Additionally, satellite brightness can be driven by a number of small specular components that are very difficult to model. The albedo-area product in the diffuse sphere model is found using a best fit to minimize rms error between observations and the model. For these observations, we find $\rho A = 0.65$. The parameters for our empirical model (without BRDF lab measurements) are similarly found using a best fit to observations. For the chassis, we find Phong parameters $K_d = 0.34$, $K_s = 0.40$, and $n = 8.9$. For the solar array, we find $K_d = 0.15$, $K_s = 0.25$, and $n = 0.26$. On the other hand, our model with BRDF lab measurements uses no prior knowledge of brightness observations. Rather, it is modeled using only the geometry and BRDF data provided by SpaceX. Figure 12 shows a comparison between the correlation of our models[7] with observations and the correlation of the diffuse sphere model.

We see that our models are roughly a 50% improvement over the diffuse sphere model, and both can be used to accurately predict satellite brightness. We notice that our model using laboratory measured BRDFs generally underpredicts brightness. This is likely due to brightness driven by satellite components we did not model in our analysis. Astronomers could account for the underestimation by fitting one free parameter (which accounts for light scatter from unmodeled surfaces) to observations. The underprediction is not an issue for satellite operators, as they are most interested in how different satellite designs increase or decrease brightness. It is important to note that this model does not incorporate any prior knowledge of observed brightness, while both our empirical model and the diffuse sphere model require many brightness observations of satellites over a range of solar angles. Additionally, we note that the diffuse sphere model only coincidentally fits the Starlink v1.5 data well, because the solar array backscatters light in a similar manner to the diffuse sphere model. Starlink V2 off-points solar arrays, so the dominant brightness is from forward scattering. A diffuse sphere model will not represent these satellites well. Our model with measured BRDFs is good enough to provide satellite operators with directional knowledge about how changing satellite design can reduce brightness. Additionally, both of our new models can be used by astronomers to estimate which areas of the night sky are least impacted by existing and future satellite constellations.

## 7. Discussion

Our motivation for this paper is to provide both satellite constellation operators and astronomers with tools for modeling satellite optical brightness. In particular, we developed a software package known as `Lumos-Sat` for satellite optical brightness predictions. In order to validate our predictions, we chose a selection of satellites for which there are time-resolved ground-based brightness observations and for which we also have BRDF data for both the satellite components and the Earth. Using SpaceX's existing Starlink v1.5 satellites, we show that our models have better predictive power than the traditional diffuse sphere calculation. For the Starlink v1.5 satellites, we find that our models are about 50% better than a calibrated diffuse sphere model. Our modeling technique can use either laboratory measured BRDFs or BRDFs that are best-fit to satellite observation data. We also note that the diffuse gray sphere model does not capture the specular nature of scattering from the satellite, in particular from the chassis deck. Despite this, the diffuse sphere model has been used in a variety of papers (Hainaut & Williams 2020; Lawler et al. 2021). This is largely because the diffuse sphere model is very simple to implement and no better brightness calculation was previously known.

[6] Typical Starlink v1.5 orbital altitude per FCC filings.

[7] Correlation does not change significantly when including earthshine, because the Pomenis data set does not include many satellite observations where earthshine dominates brightness. For simplicity, our models are shown without earthshine.

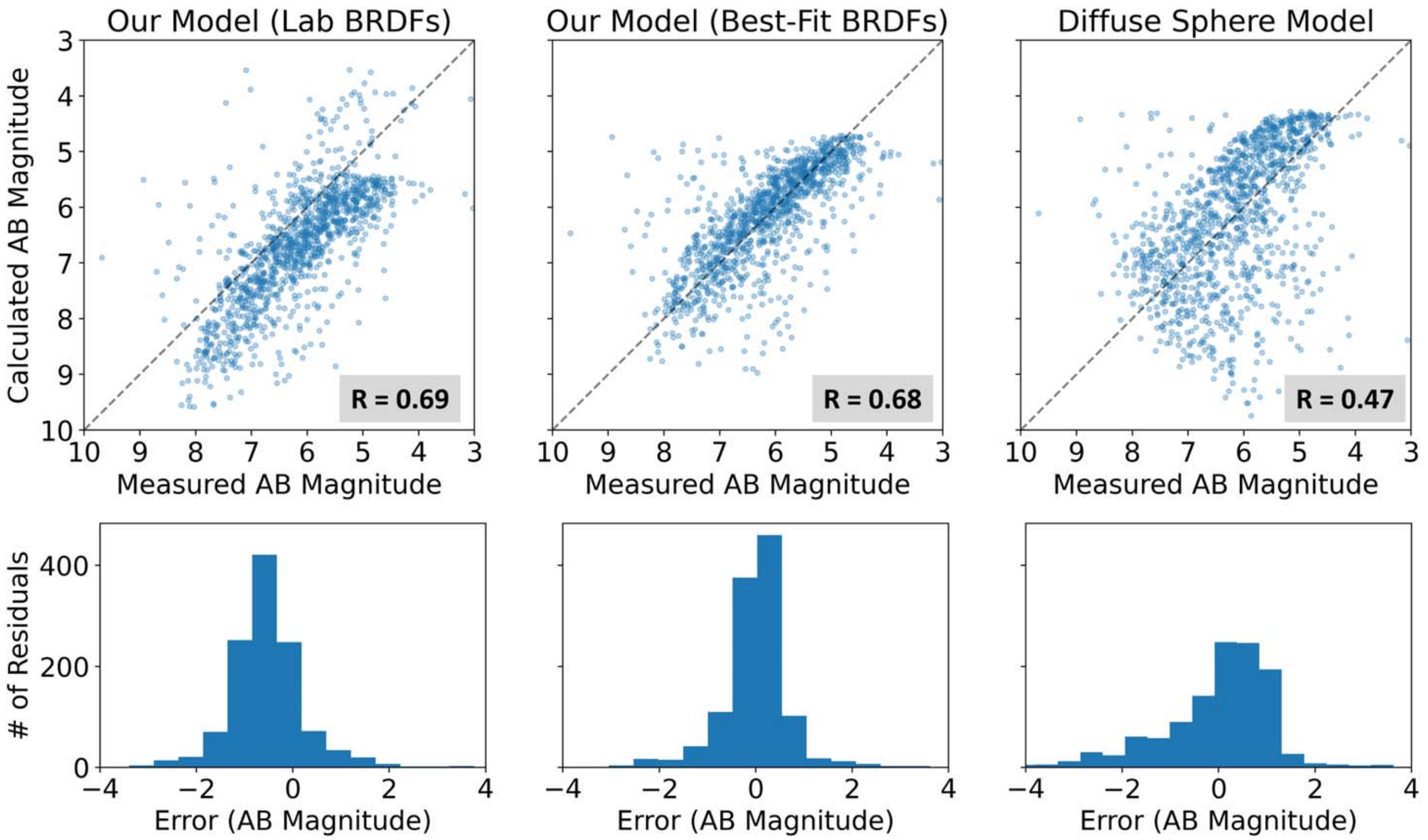

**Figure 12.** Comparison of satellite brightness models. We use Pearson correlation to measure model trend. Histograms of the residuals are also shown. For our calculations using BRDFs measured in the lab, we find a Pearson correlation of $R = 0.69$. This model systematically underpredicts brightness, because it only accounts for the two primary satellite surfaces and ignores smaller components on the satellite. For our empirical model, we find $R = 0.68$. Because this model has free parameters, it does not underpredict brightness. Both of these models are a respectable improvement over the optimal diffuse sphere model, which has a correlation of $R = 0.47$. We note that this model has systematic error, as the residual distribution is not Gaussian-like. This demonstrates that, if either lab-measured BRDFs or sufficient satellite brightness observations are available, our techniques are a better choice than the diffuse sphere model.

Improving correlation with observation is the next major challenge for this work. Moving forward, our brightness model could be improved by including mutual shadowing between satellite surfaces. For example, our current modeling does not include the effect of the chassis nadir blocking the solar array from an observer's view. This is a logical next step for our work (Cole 2020). Another shortcoming of this technique is that many satellite parameters will not be known by the public. Our model requires the normal vectors, areas, and BRDFs of the primary surfaces of a satellite. While it is known that, during nominal operation, the Starlink v1.5 chassis deck points directly nadir and that the solar array is perpendicular to the chassis deck in the brightness regime of concern, this information may not be known during orbit raising or lowering, or other off-nominal operations. However, once a BRDF model for the primary surfaces of a satellite has been found, it may be possible to use satellite brightness observations to estimate the orientations of the satellite's surfaces. This is a challenge that we leave to future research. Additionally, there is a limit to how much BRDF and geometric data will be known about satellites in the future. Generating these brightness models will require some degree of collaboration between satellite operators and astronomers. As a workaround if cooperation is not possible, BRDFs for a satellite's primary surfaces could be calibrated based on on-orbit brightness observations and limited knowledge of satellite geometry made available by FCC filings. Many LEO satellites, by necessity, will have a solar array that points toward the Sun to generate power and a chassis that points toward nadir to provide internet or other communications.

One of the most difficult parts of this analysis is accurate modeling of earthshine. It is beyond the scope of this paper, but earthshine deserves further investigation. We suggest two possible avenues for improvement. First, BRDF data for the Earth's surface could be gathered from the MODIS instrument, which operates on NASA's Terra and Aqua satellites. The MODIS instrument gathers Ross-Li BRDF coefficients for a variety of terrain (Strahler et al. 1999). Another option would be to implement analytic BRDFs, such as those used in the SHARM radiative transfer software (Lyapustin 2005).

Even when there is good collaboration with satellite operators, accounting for every small component of a satellite will be very difficult. Specularly reflecting objects on the scale of centimeters can cause "glints" (changing satellite brightness on short timescales). This includes MLI thermal blankets or even small pieces of aluminum. For these specular materials, scattered flux can change by orders of magnitude over a change in surface orientation of a few degrees or less. This issue is exacerbated by additional moving parts on a satellite bus—parabolic dishes, laser connections, or other components may rotate quickly and cause glints that are extremely difficult to predict. These glints, if not mitigated by design, are a potentially serious contributor to bogus alerts.[8] Bright glints such as iridium flares are so obvious that they would be rejected in the science analysis. Faint glints or flares are more problematic; they can imitate astrophysical flares or interfere with asteroid detection. The only mitigation is better satellite design, which can be driven by our software. Designs can be improved to reduce glints by eliminating or covering offending specular surfaces and using diffuse coatings or materials on moving components and complex geometries. While satellite glints have been detected and are a known issue (Krantz et al. 2021), little has been done to quantify or predict the impact of glints on time-domain astronomy.

[8] False detection of time-domain events by an observatory.

The ability to predict satellite brightness has major impacts for constellation operators. It allows satellite operators to use brightness as a constraint during the design process. If satellite brightness is readily predictable, operators can choose to use satellite materials and configurations or satellite conops that reduce brightness. Observations presented in this paper correspond to the first-generation Starlink satellites. While these satellites did not include brightness as a design constraint early in development, SpaceX has used brightness analysis similar to that presented in this paper to inform brightness mitigating designs on the recently developed second-generation Starlinks. Using this predictive analysis ensured that the most effective and efficient brightness mitigations were implemented on second-generation Starlinks. These mitigations include the use of specular materials on the chassis that have order-of-magnitude reductions in diffuse scatter,[9] dark paint where specular materials are not effective, and off-pointing of the solar arrays (SpaceX 2022). These improvements can be quantified using `Lumos-Sat` before satellites are in space.

Our calculations show that the sky near zenith can have bright satellite trails in astronomical twilight during dawn and dusk night observatory operations (Ivezić et al. 2019). In our simulations, earthshine causes notable satellite brightness on the eastern horizon. This possibly impacts the search for potentially hazardous asteroids (PHAs), which scans the sky toward the Sun at dawn and dusk to find asteroids interior to the earth's orbit. This works by connecting candidate detections in pairs of exposures during a night (tracklets) and extending the tracklets to additional nights (Schwamb et al. 2023). PHAs can be detected higher than 30° above the horizon. Unfortunately, earthshine can produce visible satellite trails in this region of the sky up to 40°. Bogus detections caused by satellites create noise in the tracklet extrapolation process, which causes PHA detections to fail. Our model makes it possible to quantify the impact of satellites on the PHA detection process and telescope operations more generally. Improved brightness analysis in this regime can be incorporated into PHA detection processes and satellite design and operation.

For the astronomy community, accurate brightness modeling is important to predict when and where satellites are most likely to interfere with observatory operations and data quality. Satellites have complex light scattering properties, so their optical brightness is highly dependent on their location in the sky. This knowledge can be used to inform better telescope scheduling algorithms to "dodge" LEO satellites. Satellite brightness predictions could even be used to create better satellite streak removal software. Using orbital parameters, as well as brightness prediction, the width and magnitude of a satellite streak in an image can be estimated ahead of time. Additionally, astronomers can better quantify the scientific impacts of LEO satellites.

As this is one of the first attempts to accurately predict satellite brightness, not all potential applications are known. Brightness models still need to be developed for other satellites, such as Starlink's V2 and V2 Mini, Amazon's Project Kuiper, AST's SpaceMobile, and OneWeb. We hope this promising technique and the open-source software will be further developed and utilized in the future to improve satellite design, as well as astronomy operations and data analysis.

[9] Light from the Sun that is perfectly specularly reflected by the chassis is scattered into space for all solar angles.

## Acknowledgments

We acknowledge support from NSF/AURA/LSST grant N56981CC and NSF grant AST-2205095 to UC Davis. We acknowledge useful discussions with Adam Snyder, Craig Lage, Daniel Polin, and David Goldstein. We thank Robert Blum and William O'Mullane for their careful reading of the manuscript. We thank an anonymous referee for helpful comments. We are grateful to Jared Greene, Perry Vargas, Andrew Fann, Michael Sholl, and Doug Knox for their early work on predicting Starlink brightness. We thank Tony Hobza, Veronica Rafla, and Kezhen Yin for their work in materials science that make brightness mitigations possible. We thank Harry Krantz and Eric Pearce for measurements from the Pomenis observatory.

## Data Availability

The software written for this project is available at Github: https://github.com/Forrest-Fankhauser/satellite-optical-brightness and a permanent archive can be found on Zenodo (Fankhauser 2023b). `Lumos-Sat` is an open-source Python package that was created for this project (Fankhauser 2023a). Documentation and installation instructions for `Lumos-Sat` can be found at: https://lumos-sat.readthedocs.io/. Development of `Lumos-Sat` is ongoing and collaboration is encouraged.

*Software:* Astropy (Astropy Collaboration et al. 2022), NumPy (Harris et al. 2020), SciPy (Virtanen et al. 2020), Matplotlib (Hunter 2007), pandas (McKinney 2010), SGP4, Celestrak, Space-Track.

## ORCID iDs

Forrest Fankhauser https://orcid.org/0009-0008-2880-4752
J. Anthony Tyson https://orcid.org/0000-0002-9242-8797
Jacob Askari https://orcid.org/0009-0008-3846-9708

## References

Astropy Collaboration, Price-Whelan, A. M., Lim, P. L., et al. 2022, Apj, 935, 167
Barentine, J. C., Venkatesan, A., Heim, J., et al. 2023, NatAs, 7, 252
Cole, R. E. 2020, RNAAS, 4, 182
Elvidge, C. D., Baugh, K., Zhizhin, M., Hsu, F. C., & Ghosh, T. 2017, IJRS, 38, 5860
Fankhauser, F. 2023a, Lumos-Sat, v1.0.7, Zenodo, doi:10.5281/zenodo.8048257
Fankhauser, F. 2023b, Satellite Optical Brightness Software, v1.0.0, Zenodo, doi:10.5281/zenodo.8048219
Gatebe, C. K., & King, M. D. 2016, RSEnv, 179, 131
Germer, T. A., & Asmail, C. C. 1997, Proc. SPIE, 3141, 220
Greynolds, A. W. 2015, Proc. SPIE, 9577, 95770A
Hainaut, O. R., & Williams, A. P. 2020, A&A, 636, A121
Harris, C. R., Millman, K. J., van der Walt, S. J., et al. 2020, Natur, 585, 357
Hu, J. A., Rawls, M. L., Yoachim, P., & Ivezić, Ž. 2022, ApJL, 941, L15
Hunter, J. D. 2007, CSE, 9, 90
Ivezić, Ž., Kahn, S. M., Tyson, J. A., et al. 2019, ApJ, 873, 111
Krantz, H., Pearce, E. C., & Block, A. 2021, arXiv:2110.10578

Lawler, S. M., Boley, A. C., & Rein, H. 2021, AJ, 163, 21
Lawrence, A., Rawls, M. L., Jah, M., et al. 2022, NatAs, 6, 428
Lyapustin, A. 2005, ApOpt, 44, 7764
Matusik, W., Pfister, H., Brand, M., & McMillan, L. 2003, ACM Trans. Graph., 22, 759
McKinney, W. 2010, in Proc. IX Python in Science Conference, ed. S van der Walt & J Millman (Austin, TX: SciPy), 56
Nattinger, K. 2020, Proc. SPIE, 11485, 1148503
Pearce, E. C., Avner, L., Block, A., Krantz, H., & Rockowitz, K. 2018, in The Advanced Maui Optical and Space Surveillance Technologies Conf, ed. S. Ryan (Kihei, HI: AMOS), 56
Phong, B.-T. 1973, PhD thesis, The University of Utah
Schwamb, M. E., Jones, R. L., Volk, K., et al. 2023, ApJS, 266, 22
SpaceX 2022, Brightness Mitigation Best Practices for Satellite Operators, https://api.starlink.com/public-files/BrightnessMitigationBestPracticesSatelliteOperators.pdf
Strahler, A. H., Muller, J., Lucht, W., et al. 1999, Algorithm Theoretical Basis Document v5.0, MODIS BRDF/Albedo Product, https://modis.gsfc.nasa.gov/data/atbd/atbd_mod09.pdf
Tyson, J. A., Ivezić, Ž., Bradshaw, A., et al. 2020, AJ, 160, 226
Venkatesan, A., Lowenthal, J., Prem, P., & Vidaurri, M. 2020, NatAs, 4, 1043
Virtanen, P., Gommers, R., Oliphant, T. E., et al. 2020, NatMe, 17, 261
Wanner, W., Li, X., & Strahler, A. H. 1995, JGRD, 100, 21077